\documentclass[
 ,draft            
  ]
  {aipproc}

\layoutstyle{6x9}

\begin{document}

\title{The case for an aggressive program of dark energy
  probes\footnote{Invited talk at PASCOS 07, Imperial College, July 2007}}


\author{Andreas Albrecht}{
  address={Department of Physics, One Shields Ave.; University of
  California; Davis, CA 95616}
}

\begin{abstract}
The observed cosmic acceleration presents the physics and cosmology
communities with amazing opportunities to make exciting, probably even
radical advances in these fields.  This topic is highly data driven
and many of our opportunities depend on us undertaking an ambitious
observational program. Here I outline the case for such a program
based on both the exciting science related to the cosmic acceleration and
the impressive impact that a strong observational program would
have. Along the way, I challenge a number of arguments that skeptics
use to question the value of a strong observational commitment to
this field.  
\end{abstract}

\maketitle


\section{Introduction}

This is truly remarkable time to be involved in cosmology
research.  There are many reasons for this, but none stand out quite
as dramatically as the observed cosmic acceleration (attributed in
current nomenclature to the ``dark energy'').  In the words of
the Dark Energy Task Force (DETF)~\cite{Albrecht:2006um} ``most experts believe
that nothing short of a revolution in our understanding of fundamental
physics will be required to achieve a full understanding of the cosmic
acceleration''. As things stand, this revolution is being motivated both
by remarkable new data sets and by exciting theoretical developments. 

Many ambitious researchers in cosmology and related fields have been
galvanized by these extraordinary developments.  There have
been a number of excellent talks at this conference on the topic of
cosmic acceleration which capture some of this climate.  The DETF, charged
with charting a way forward with dark energy observations, received
50 thoughtful and thorough whitepapers from leaders in the field
(despite the lack of any specific commitment at that time to fund
future dark energy experiments).

Most of us look at these developments and are astonished at our good
fortune to be part of what will surely be viewed as one of the great
moments in the history of science.  Indeed, since the discovery of dark
energy every group that has deliberated on future directions for the
field has recognized the exciting opportunities and challenges
presented by the cosmic acceleration\footnote{See for example
  \cite{Turner:2003pe,Albrecht:2005np,Peacock:2006kj}. Since the 2006 DETF report a 
  number of panels have recommended pursuit of specific ground and
  space based dark energy projects, and just since PASCOS 07 the US
  National Research Council's {\em Committee on NASA's Beyond Einstein
  Program} named a ``Joint Dark Energy Mission'' the top
  priority for that program
  \url{http://nationalacademies.org/morenews/20070907b}}. 
But a small number of negative voices continue to be heard
alongside the building enthusiasm for further studies of the dark
energy.  To some degree this negativity may reflect the natural
skepticism of scientists in the face of unbridled enthusiasm (such as
the above).  To the extent that that is the explanation, the
skepticism is surely a good thing that will lead to a healthy debate
and produce more rigorous research.  On the other hand, I suspect some
of the negativity is just the result of sloppy thinking and needs to
be challenged and simply put to rest.  Regardless of how one might
attribute explanations and motivations, one purpose of this paper is to
engage in this debate on a number of fronts.  

Here are some illustrations of some of the negative views I am talking
about. At lunch on the first day of the PASCOS 07 conference there was
a lively discussion about the merits of dark energy experiments.  One
 colleague commented 
\begin{quote}
  {\em ``Studies of dark energy are unlikely to be interesting because we
  already have a theory of dark energy.''}
\end{quote}
moments later, another  cosmologist declared
\begin{quote}
  { \em ``Studies of dark energy are unlikely to be interesting because we
  have no theory of dark energy.''}
\end{quote}
The two were united in their conclusion, and apparently not too bothered
by subtle differences in their reasoning

One unavoidable feature of physics research is that at any given time
the total amount of data is finite.  That necessarily means there will
be more than one theory that fits the data.  In many fields, this
universal fact is seen as a source of vitality. Curiosity about
further resolving these degeneracies drives exciting new experiments
and theoretical work. The momentous progress in physics over the last
century can be seen in this light as the fundamental particles, the
atomic theory of heat, quantum physics and general relativity emerged
from ``under the radar'' of earlier data sets that were fit perfectly
by more primitive theories.  As was amply evident in the talks at the
PASCOS 07 conference, we have good reason to look forward to similar
advances as the LHC data starts coming in. 

But for some reason, the fact that a given future dark energy
experiment will not remove all uncertainties about the nature of dark
energy seems to generate considerable angst among some physicists and
astronomers\footnote{See the question section of my PASCOS 07 talk
for an example\cite{PASCOS07}} and is sometimes given as a reason
to be discouraged from even doing more experiments. As I shall
quantify below (and as was also shown by others at PASCOS 07), the
proposed new experiments will have an impressive impact on our
knowledge of dark energy.  This is all one can ever ask of a new
experiment. 

This paper focuses on two key areas.  In the next section I
review some of the thriving theoretical work that has been stimulated by the
cosmic acceleration. The striking theoretical issues raised by the
cosmic acceleration and the remarkable directions we have been driven
in our initial attempts to understand it are the key reasons I find
this topic so deeply interesting. I also believe this is why so many
excellent researchers are taking risks and changing direction in their
careers in order to get involved. 

The third section summarizes a number of results which
demonstrate the tremendous impact future experiments can have on our
understanding of dark energy, both on our understanding of the general
properties of dark energy and in terms of constraining specific models
of dark energy that are currently of interest.  It is these results
that demonstrate that an ``aggressive program of dark energy probes''
is indeed possible. 

This paper is based on my talk at the PASCOS 07 meeting at Imperial
College.  My slides and a video of the talk are available
online\cite{PASCOS07}. This online material as well as my ``Origins of
Dark Energy'' talk\cite{ODE} and related papers\cite{Albrecht:2007qy,Abrahamse:2007ip} are a good
source for the technical material on which this paper is based.  My
goal here is to assemble some key arguments in a concise form.
Readers seeking more details should refer to this other material. 

\section{Dark Energy Science}
\label{Sect:DES}

\subsubsection{Two types of acceleration}
At the start of the era of observations of cosmic acceleration the
cosmology community was familiar with two possible sources of cosmic
acceleration. These two types of acceleration were sufficiently
established that they could both be found in textbooks. 

{\bf Cosmological constant:} The cosmological constant arises as an
``extra'' constant term in 
Einstein's equations. A FRW universe with a cosmological constant with
value $\Lambda$ is equivalent to adding a matter component with
density $\rho_\Lambda \equiv {\Lambda \over 8\pi G}$ and with equation
of state $p =-\rho$. This equation of state insures that
$\rho_\Lambda$ remains constant throughout the evolution of the
universe.  Until the mid 1990's, a commonly held belief among particle
physicists and cosmologists was that $\Lambda=0$.  For many this
belief was partly motivated by the fact that naively quantum fields
give contributions to $\Lambda$ of order $10^{60}$ or $10^{120}$ times
larger than observationally acceptable values.  It was widely believed
that finding some symmetry or dynamical process that sets $\Lambda$
(including contributions from quantum fields) precisely to zero was
the best hope to resolve this apparent discrepancy.  

{\bf Dynamical acceleration:} Another widely held belief that has had
growing support over the last few decades is that the universe
underwent a period of {\bf cosmic inflation} in the distant
past.  A period of cosmic inflation  appears to explain many
features observed in the universe today, some of which 
seems puzzling before the idea of inflation came along.  

Today inflation is quite well understood in terms of its
phenomenology, but it still has a number of unresolved foundational
questions. Despite these, it is certainly clear that cosmic inflation
requires a period of cosmic acceleration that {\em cannot} be
described by a cosmological constant.  Cosmic inflation is understood
to be driven by some matter field typically called the ``inflaton''
which exhibits an equation of state $p = w\rho$.  During inflation
$w$ takes values that approach $w=-1$ but it is a requirement of
inflation that strict equality does not hold.  This is because
inflation is fundamentally a dynamical process. Not only must the
universe enter and exit inflation (leading to large variations in $w$)
but small deviations from $w=-1$ are also necessary throughout the
inflationary period in order for the mechanisms of inflation to work
properly. 

Today, some people (one quoted in the introduction) like to push the
point of view that there is one superior theory of cosmic acceleration,
namely the cosmological constant.  It is very likely that before the late
1990's, these same people believed in cosmic inflation, and also
believed $\Lambda=0$.  In fact, the pioneers who started contemplating
$\Lambda \neq 0$ in the 1990's \cite{Efstathiou:1990xe,Krauss:1995yb}
did so precisely to prop up inflation (and the associated ``CDM''
cosmology) in light of data which would have otherwise been problematic.
In fact \cite{Efstathiou:1990xe} notes the possibility of a dynamical source of
acceleration and even seems to regard it somewhat favorably over a
strict cosmological constant.   

So the field really has two different ideas of how cosmic
acceleration can come about.  One is dynamical (inflation-like) and
one is not dynamical (the cosmological constant).   Because of this 
interesting dichotomy the DETF (and many others) see the
discrimination between dynamical and non-dynamical sources of cosmic
acceleration as the best starting point for analyzing dark energy data
(both present and future). 

Note that {\em if} the data continues to be consistent with a cosmological
constant it will only ever {\em bound} the possibility that there is a
dynamical aspect to the cosmic acceleration. In that case one can
never prove that 
the acceleration is absolutely constant.  This should be no more
troubling to a physicist than the fact that we only have {\em bounds}
on the photon mass, deviations from the equivalence principle and 
other aspects of the physical world that many presently regard as
absolute truths.  The reason  massless photons and the equivalence
principle are seen as ``truth'' is that we have sufficiently high
quality data to know that any corrections will be very small.  The
impossibility of getting an absolute result in a cosmological constant
driven universe should certainly not make us timid about seeking high
quality data in this area as well. 

\subsubsection{Some remarkable features of accelerating cosmologies}

A key part of the science case for further study of dark energy
is the very exciting nature of the theoretical developments that surround
this this topic.  I will highlight a few of these in this section. 

{\bf Strange numbers:}
I've already mentioned the fact that naive quantum field theory arguments
predict a discrepancy in the value of $\Lambda$ by a factor of $10^{60}$
or even $10^{120}$.  In addition, most dynamical models of dark energy
end up requiring a new particle associated with a so called
``quintessence field'' that has a mass of $10^{-31}eV$ (many orders of
magnitude smaller than current bounds of the photon mass).  With a few
noteworthy exceptions, theories of quintessence do not consider the
problem of how to protect such a tiny mass from quantum corrections.
Also, in general such a light particle will lead to long range forces,
and most quintessence models do not have a mechanism for evading
current bounds on such forces.  Some dynamical models require
additional tuning of parameters so that the onset of cosmic
acceleration can occur in the right epoch of the universe (the ``why
now problem'').  Theorists tend to take such features of a theory as a
sign that something important is missing in our understanding.

{\bf $\Lambda$ and equilibrium:}
If the universe has a true (non dynamical) cosmological constant with
a positive value (sufficient to account for the current acceleration, or
even smaller) the future of the universe will have some striking
features.  The effective ``cosmological constant density''
$\rho_\Lambda$ will come to completely dominate the universe as the
other types of matter dilute with the cosmic expansion.  Under these
conditions the universe will approach something called ``de Sitter
space'', which has features which fit a (somewhat generalized) notion
of equilibrium:  The de Sitter space is approached asymptotically by
huge fraction of all possible initial states, it has a (Hawking)
temperature and an entropy.  The entropy of a de Sitter space has been
shown to be larger than the entropy of all other cosmological states
with a non-zero cosmological constant\cite{Gibbons:1977mu}. 

This situation implies the following picture of cosmology in the
presence of a fundamental cosmological constant: The universe is
eternal, and spends most of its time in the equilibrium state
described above.  Cosmology as we know it must emerge from very rare
large fluctuations from equilibrium.  I have argued
elsewhere\cite{Albrecht:2004ke} that this general picture of cosmology is much
more powerful than more traditional approaches to cosmological
``initial conditions''.  This is because the probabilities are in
principle assigned to different cosmologies (or fluctuations from
equilibrium) by concrete calculations based on fundamental physical
laws rather than ad hoc proposals for ``wavefunctions of the universe''
which, far from being compelling, are notorious for generating
controversy. 

However, the first authors to calculate probabilities in this
equilibrium picture concluded that that picture is a spectacular
failure because cosmologies inconsistent with our observations were
exponentially favored by their calculations\cite{Dyson:2002pf}.  Sorbo and
I\cite{Albrecht:2004ke} have proposed an alternative approach to these
calculations which exponentially favors the standard
inflationary picture of cosmology. Deeper insights into the nature of quantum gravity
are needed to learn which (if either) of the two approaches is
correct.  

My main point here is that considering the cosmic acceleration has
driven theorists to contemplate radical new ideas about how to think
about cosmology, equilibrium and initial conditions. 

{\bf The string theory landscape:} 
Andrei Linde and Renata Kallosh have given excellent talks at
PASCOS 07 on the string theory landscape.  I will refer you to those
for details.  I will just make a few brief remarks here on topics that
link the string theory landscape to the main focus of my discussion. 

The string theory landscape marks a radical change in thinking about
fundamental physics.  For many, it has replaced the old idea that the
correct ``theory of everything'' will state the fundamental particles
and their parameters and couplings in one tidy package.  Instead, the
string theory landscape tells us that when it comes to the physical
observables that are important to us, they are chosen form an enormous
collection (around $10^{1000}$) of string theory ``vacuua'' that
fundamentally appear to be equally valid choices. 

As Linde and Kallosh have emphasized, this dramatically different
approach to fundamental physics was driven in large part by attempts to reconcile string
theory with the observed cosmic acceleration. In fact, many people
believe that this picture leads to a successful
(anthropic\cite{Weinberg:1988cp}) account of 
how the cosmological constant (or more correctly, the vacuum energy)
should have more or less the value needed to explain the current
cosmic acceleration\cite{Bousso:2000xa}. Some of the most enthusiastic
proponents of this view have even argued informally that this success
is so compelling as to undermine the need to collect further data on
the cosmic acceleration.  

However, as you heard in other talks at this meeting, there is more to the
landscape vacuua than stating that they all fundamentally appear to be
equally valid.  The dynamics of a state evolving in the landscape will
tend to prefer some vacuua over others.  Perhaps even an equilibrium
picture could emerge. Basically, one has to do full cosmology in the
landscape before real predictions can be made.  I find the work along
these lines extremely interesting, but the landscape is a complicated
place and we appear to be very far from a final answer.  It is
entirely possible that once we understand the full dynamics of the
landscape the results will be inconsistent with our observations (just as was
found in \cite{Dyson:2002pf} for the $\Lambda$ case),
causing a complete failure of the landscape picture. Also, the
acceleration in the landscape picture is expected to be
observationally indistinguishable from a cosmological constant, so any
observation signal of $w\neq -1$ will rule out the landscape explanation.

The string landscape is another example where attempts to understand the cosmic
acceleration have driven theory in dramatic new directions.

{\bf Other explanations:} 
There are a variety of other approaches to explain cosmic acceleration.
These include modifying Einstein gravity (see \cite{ODE} for numerous
talks on this subject), introducing new non-accelerating physics to
explain the observations\cite{Csaki:2001yk,Csaki:2004ha}, or ``even''
considering the possibility that more conventional physics produce
acceleration \cite{Deser:2007jk} or lead
to observations that we are currently misinterpreting as
acceleration\cite{Kolb:2005da,Martineau:2005zu}. 

\section{The impact of future experiments}
\label{Sect:TIOF}

\subsubsection{The DETF}

Based on 50 whitepapers, as well as internal expertise, the DETF
developed mock data (or ``data models'') to represent a number of
future dark energy observations.  They divided these data models into
stages.  ``Stage 2'' represents data from projects that are already
underway, ``Stage 3'' projects are medium term medium cost projects,
and ``Stage 4'' are the larger more expensive projects currently under
consideration. 

The DETF considered the impact of the various projected data sets on
cosmological parameters in a standard cosmological model. They
parameterized the dark energy by its density today as well as a two
parameter description of the equation of state parameter as it evolves
with the cosmic scale factor $a$ given by $w(a) = w_0 + w_a(1-a)$ (the
``$w_0-w_a$'' parameterization).    

The DETF considered cosmologies in which the cosmic curvature is
allowed to be nonzero ($\Omega_k \neq 0$). While some are happy to
take $\Omega_k=0$ as a prior assumption, I don't believe there is a good
case for doing so.  For one, the main reason people like the
$\Omega_k=0$ prior is that that value is seen as a universal prediction of
cosmic inflation.  However, the only reason the $\Omega_k=0$ universe
has any chance of fitting the data is due to the presence of dark
energy.  Since the dark energy is so poorly understood, I feel it is a
mistake to assume we know how to include it correctly, and how well it
may or may not fit the $\Omega_k=0$ picture when things are understood
more completely.  Secondly, there is a very interesting subclass of
inflation models that allow a small deviation from
$\Omega_k=0$ (see for example \cite{Freivogel:2005vv}). Discriminatory power between these pictures is a valuable
capability to expect from future observations. 

The DETF used a figure of merit given by the inverse area in $w_o-w_a$
space constrained by a given experiment.  The DETF showed that good Stage 3
data will improve on Stage 2 constraints in $w_0-w_a$ space by a factor of order
3, and good Stage 4 experiments will result in an order of magnitude
improvement. We note that in the parameter space considered by the DETF
the current data (Stage 1) does not result in significant constraints,
which is why current data is typically  analyzed in terms of simpler
cosmological models\cite{Knox:2005hx}.

\subsubsection{Beyond the DETF}
The DETF report gives a useful perspective on the power of future
experiments, but it left a number of interesting answered
questions. One question is ``did the DETF miss anything important by
using their particular parameterization of the dark energy?''  Some
researchers are also concerned that the DETF dark energy
parameterization is too abstract, and would like to see estimates of
the impact of future experiments on specific existing models of dark
energy.  In the next two sections I will report on new work (since the
DETF) that answers both these questions, and helps give a more
complete picture of the full power of proposed dark energy
experiments. Another key ``beyond DETF'' question involves the
subtleties of combining data sets.  This is an area where improvements
on the DETF analysis will certainly be important and new
work\cite{Zhan:2006gi,Schneider:2006br} suggests such improvements
will demonstrate an even greater 
impact for future experiments. Those improvements are not included in
the analysis below, so these results should be seen as underestimates
of the true potential of future experiments. 

{\bf Beyond two parameters:} 
The $w_0-w_a$ parameterization used by the DETF allows a variety of
linear functions $w(a)$.  With Bernstein, I modeled $w(a)$ with a
much larger number of parameters\cite{Albrecht:2007qy}, resulting in
stepwise constant functions $w(a)$. We used a Fisher matrix analysis that
allowed us to consider an orthonormal basis of independently measured mode
functions $w_j(a)$.  An advantage of this approach is that when one
increases the total number of parameters (by decreasing the step width)
one simply gets an ever improving approximation to the finite number of
continuous modes that are well-measured by the given experiment (most other
parameterizations of $w(a)$ are highly unstable against changing the
number of parameters).  We argue in \cite{Albrecht:2007qy} that in the
small bin limit our approach allows each experiment to show exactly
what it is able to measure.  The well-measured modes are chosen by the
properties of the experiment and thus there is no information lost due
to prior assumptions about the form of $w(a)$.  

\begin{figure}
\includegraphics[height=.4\textheight]{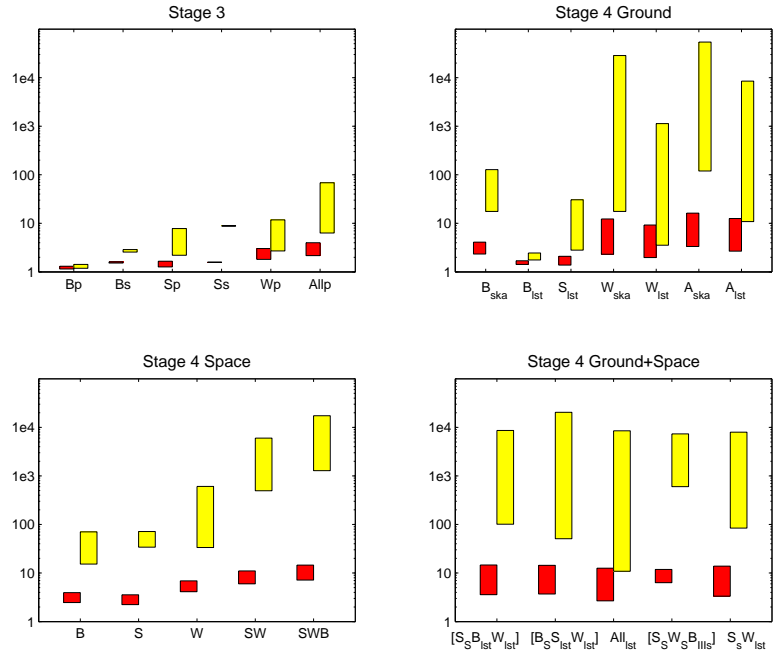}
  \caption{\label{bars} Figure 3 from \cite{Albrecht:2007qy} showing the DETF
  figure of merit (dark bars) and the figure of merit given by a
  complete set of well measured modes $w_j(a)$ (light bars).  As
  detailed in \cite{Albrecht:2007qy}, the 
  higher dimensional parameter space gives a more complete picture of
  the ability of a given experiment to constrain dark energy
  properties.  Good stage 4 experiments have a
  figure of merit many orders of magnitude higher than the DETF
  estimates. } 
\end{figure}

Figure \ref{bars} shows the figure of merit derived using DETF
parameters (dark bars) and a suitably large number of step parameters
to achieve convergence onto the well-measured modes (light bars).  
The different bars and panels correspond to different choices of DETF data
models\cite{Albrecht:2007qy}.  The upshot of \cite{Albrecht:2007qy} is that the best future
data sets will measure many more than two parameters, resulting in
massively higher figures of merit vs. the DETF estimates. These huge
figures of merit translate directly into greater discovery power through
greatly improved constraints in parameter space.  However,
aside from an overall rescaling of the figure of merit, the other
conclusions of the DETF regarding the importance of combining
techniques and the relative rankings of the different data models are
unchanged.  

{\bf Beyond abstract parameters:} 
Theories of dark energy appear to be in a very primitive state, and
are likely to change greatly before we have a deep understanding of
the cosmic acceleration.  To some the use of abstract parameters such
as I've used above is the best approach. Basic characteristics
of the dark energy can be discussed (such as whether or not it is
dynamical) without dependence on a particular model that is unlikely
to survive progress in this field.

Others would prefer to know if future experiments will have an impact
on existing proposed explanations of dark energy.  However flawed
the existing explanations may turn out to be, they are what we have to
work with for now and one certainly should expect proposed experiments
to have an impact on these models.

To assess this point, my students and I have undertaken a number of
studies of the impact of DETF data models on specific models of dark
energy.  You can learn more about this work in
\cite{PASCOS07,ODE,Abrahamse:2007ip}. Here I will highlight a couple
of main results.   

Firstly, we found that the parameters of the specific models
(i.e. parameters in the quintessence potential) were constrained to a
similar degree as the DETF parameters.  Typically a given model had two
parameters that were constrained by cosmology, so the comparison to the
two parameter DETF scheme made sense.  In this picture, the higher
number of parameters found in \cite{Albrecht:2007qy} shows up as
discriminating power between a large variety of dark energy models
(even though each tends to have a smaller number of parameters,
parameterizing a limited set of functions $w(a)$ that are specific to
that model).   t

\begin{figure}
  \includegraphics[height=.3\textheight]{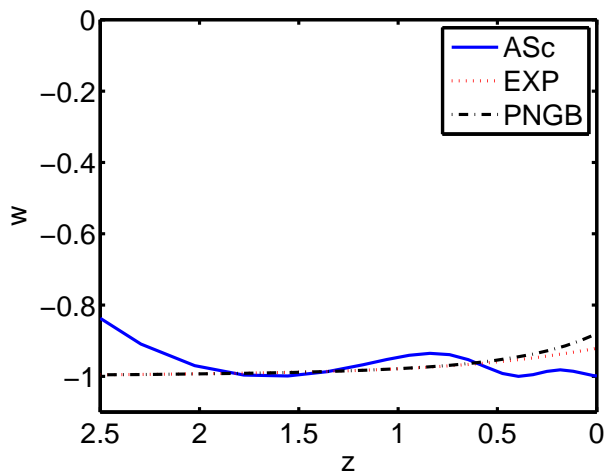}
  \caption{\label{all} The three curves give $w(a)$ for three
  different models of dark energy.  If the universe has chosen any of
  these models Stage 4 data would exclude a cosmological constant at
  at least {\em four sigma }. }
\end{figure}

Figure \ref{all} gives another illustration  of the impact of future
experiments. 
 The function $w(a)$ is shown for three different
quintessence models with parameters fixed at specific values.  If the
universe has chosen any of these three models Stage 4 data will exclude
a cosmological constant to at least four sigma. For these, the
discriminatory power of even the best Stage 3 experiments would be
below two sigma.  

\section{Conclusions and summary}
\label{Sect:CAS}

The case for aggressive pursuit of new data on dark energy is twofold.
Firstly, I have outlined how the subject of dark energy has generated
very exciting and often radical new theoretical ideas.  These include
dramatic proposals that change how we think about equilibrium and
initial conditions in cosmology and even how we formulate fundamental
theories. Second, impressive new experiments are within reach that
could have a tremendous impact on our understanding of dark energy.
The best experiments will constrain dark energy properties orders of
magnitude better than the current or medium sized future experiments,
and will have the ability to strongly discriminate among and even
fully eliminate popular dark energy models based on subtle variations
in the equation of state. 

I have outlined and challenged some of the skeptical perspectives I have heard
regarding future dark energy studies. The discovery of the cosmic
acceleration has caused a great upheaval in our thinking about
fundamental physics and cosmology.  I often sense that the skeptics
are hoping this upheaval will end quickly and are grasping for
arguments that will allow things to rapidly return to
normal.  I feel this outcome is very unlikely, and this is exactly why
I find the topic so exciting. 

Nature has handed us an amazing opportunity. I hope that the physics
and cosmology communities have the strength to face the challenge of
the cosmic acceleration head on and give a response that we can be
proud of when people write the history of this era.


\begin{theacknowledgments}
  I would like to thank A. Abrahamse, M. Barnard,
  G. Bernstein, B. Bozek, L. Sorbo, M. Yashar and the DETF members who
  collaborated with me on some of the work reviewed here, and
  B. Bozek for helpful comments on the manuscript. Also, I thank the organizers,
  especially Arttu Rajantie, for a really excellent conference.  This
  work was supported in part by DOE grant DE-FG03-91ER40674 and NSF
  grant AST-0632901.  
\end{theacknowledgments}



\bibliographystyle{aipproc}   

\bibliography{AAPSCVB}

\IfFileExists{\jobname.bbl}{}
 {\typeout{}
  \typeout{******************************************}
  \typeout{** Please run "bibtex \jobname" to optain}
  \typeout{** the bibliography and then re-run LaTeX}
  \typeout{** twice to fix the references!}
  \typeout{******************************************}
  \typeout{}
 }

\end{document}